\documentclass{PoS}

\usepackage{amsmath}
\usepackage{braket}

\renewcommand{\vec}[1]{\mathbf{#1}}

\title{Pion-pion scattering and the timelike pion form factor 
	from $N_{\mathrm{f}} = 2+1$ lattice QCD simulations using the 
	stochastic LapH method}

	\ShortTitle{Pion-pion scattering and the timelike pion form factor} 

	\author{John Bulava and  
	Ben H\"{o}rz\thanks{Combined proceedings from the talks of J. Bulava and B. H\"{o}rz.}\\
        School of Mathematics, Trinity College Dublin\\
			  Dublin 2, Ireland \\
				E-mail:\email{jbulava@maths.tcd.ie}, \email{hoerz@maths.tcd.ie}}

				\author{Brendan Fahy\\
					High Energy Accelerator Research Organization (KEK)\\
				  Ibaraki 305-0801, Japan\\
					E-mail:\email{bfahy@post.kek.jp}}

				\author{K. J. Juge\\
				Dept. of Physics, University of the Pacific\\
				Stockton, CA 95211, USA\\
				E-mail:\email{kjuge@pacific.edu}}

				\author{Colin Morningstar\\
        Dept. of Physics, Carnegie Mellon University\\
				Pittsburgh, PA 15213, USA\\
				E-mail: \email{colin\_morningstar@cmu.edu}}

				\author{Chik Him Wong\\
				Dept. of Physics, University of Wuppertal\\
				Gaussstrasse 20, D-42119 Germany\\
				E-mail: \email{cwong@uni-wuppertal.de}}

\abstract{We report on progress applying the stochastic LapH method to
estimate all-to-all propagators required in correlation functions of
multi-hadron operators relevant for pion-pion scattering. Large-volume results 
for $I=2$ and $I=1$ pion-pion scattering phase shifts 
with good statistical 
precision are obtained from an $N_{\rm f} = 2+1$ anisotropic Wilson clover ensemble 
with 
$m_{\pi} = 240\mathrm{MeV}$. We also present a preliminary 
determination of 
the $I=1$ pion-pion scattering phase shift and timelike 
pion form factor on an isotropic $N_{\rm f}=2+1$ flavour ensemble generated by the 
Coordinated Lattice Simulation (CLS) community effort.}

\FullConference{The 33rd International Symposium on Lattice Field Theory\\
                 14 -18 July  2015\\
                 Kobe International Conference Center, Kobe, Japan}

\begin{document}

   Lattice QCD simulations are inevitably carried out in finite volume and 
	 euclidean time, 
	which complicates scattering calculations~\cite{Maiani:1990ca}. 
	Since most excited hadrons 
	are unstable resonances which appear experimentally as features in 
	scattering cross sections, first-principles calculation of hadron-hadron 
	scattering amplitudes is desirable. 
  The relation between finite-volume two-hadron spectra 
	and infinite-volume elastic scattering amplitudes was 
	formulated by L\"{u}scher~\cite{Luscher:1990ux,Luscher:1991cf} more than two 
	decades ago and extended to moving frames in Ref.~\cite{Rummukainen:1995vs}. 
	However, only recently have algorithmic advances in lattice QCD spectroscopy 
	enabled finite-volume spectra to be calculated efficiently in large volume 
	with light pion masses. 

  These advances concern the treatment of all-to-all propagators, 
	which are required to give definite momenta to all hadrons and  
	to treat valence-quark-line-disconnected Wick contractions. 
	Laplacian-Heaviside (LapH) quark 
	smearing projects the quark propagator onto the subspace 
	span\-ned by the lowest-lying $N_{v}$ eigenmodes of the three-dimensional 
	covariant Laplace operator~\cite{Peardon:2009gh}. Stochastic noise introduced only in the LapH subspace results in more efficient 
	estimators for  
	all-to-all quark propagators compared to noise on the entire lattice~\cite{Morningstar:2011ka}. 

	The spatial profile of the LapH subspace projector is approximately Gaussian, as with other quark smearing procedures. In order to maintain a constant 
	physical smearing radius, $N_{v}$ must increase proportionally to the 
	spatial volume. However, in Ref.~\cite{Morningstar:2011ka} it was demonstrated that with a moderate amount of dilution in the LapH subspace, the number of 
	quark matrix inversions can be held constant as the spatial volume is 
	increased without increasing the stochastic estimation error relative to the 
	gauge noise.
	This enables all-to-all quark propagators to be estimated efficiently in 
	large spatial volumes for a reasonable cost.

	The efficient treatment of all-to-all quark propagators in turn enables 
	precision calculation of correlation functions containing multi-hadron 
	interpolating operators with definite momenta and/or disconnected Wick 
	contractions. From these correlation functions, finite-volume energies 
	can be precisely extracted, which then yield elastic scattering 
	amplitudes. The application of these techniques to extract elastic pion-pion 
	scattering amplitudes from large volume ensembles is the subject of this 
	work.
 Sec.~\ref{s:aniso} details a first application of the stochastic LapH method 
 to large volume, in which 
	the $I=1$ and $I=2$ elastic pion-pion scattering phase shifts are calculated. 
	With an eye toward larger, finer lattices at lighter pion masses, 
	Sec.~\ref{s:ff} presents a preliminary calculation of the $I=1$ 
	scattering phase shift and timelike pion form factor on an ensemble 
	generated through the Coordinated Lattice Simulations (CLS) community 
	effort. 

	\section{$I=1$ and $I=2$ 
	phase shifts from a large-volume anisotropic lattice}\label{s:aniso} 

 	For a first large-volume application of the stochastic LapH method, 
	we employ an anisotropic lattice regularization, in which the 
	spatial lattice spacing ($a_s$) is larger than the temporal one ($a_t$). 
	The renormalized anisotropy $\xi_R = \frac{a_s}{a_t}$ 
	is defined by demanding that pions satisfy the correct (continuum) dispersion
	relation
	\begin{align}\label{e:disp}
(a_{t}E_{\pi})^2 = (a_tm_{\pi})^2 + \left(\frac{2\pi a_s}{\xi_R L}\right)^2 \vec{d}^2,
	\end{align}
	where $\vec{d} \in \mathbb{Z}^3$ is the 
quantized finite-volume momentum. Details on the anisotropic ensemble used 
here are found in Tab.~\ref{t:aens}. 
\begin{table}
\centering
	\begin{tabular}{|c|c|c|c|c|c|}
		\hline
		$(L/a_{s})^3\times(T/a_{t})$ & $m_{\pi} (\mathrm{MeV})$ & $a_t (\mathrm{fm})$ & 
		$\xi_R$ & $m_{\pi}L$ & $N_{\mathrm{cfg}}$ \\
		\hline
		$32^3 \times 256$ & $240$ & $0.035$ & $3.4418(94)$ & $4.3$ & $412$ \\ 
		\hline
	\end{tabular}
	\caption{\label{t:aens}Details for the anisotropic $N_f = 2+1$ gauge 
		configurations used here.  For a complete discussion of the 
		ensemble generation and 
		scale setting,
	see Ref.~\cite{Lin:2008pr}. For definition of the LapH subspace and 
specification of dilution schemes, see Ref.~\cite{Morningstar:2011ka}.}
\end{table}

The anisotropy is crucial in defining the 
center-of-mass momentum (discussed later), and thus care must be taken in 
its determination. We employ two different methods 
which are consistent within statistical errors. In the first  
we calculate $a_tE_{\pi}$ for all pions with $\vec{d}^2 \le 5$ on $N_b = 800$ 
identical bootstrap samples of the corresponding correlation functions 
using correlated 
$\chi^2$ fits. We then perform correlated $\chi^2$ fits on each bootstrap 
sample to Eq.~\ref{e:disp} to obtain $\xi_R$. Our second determination of 
$\xi_R$ simultaneously fits all of these correlation functions to obtain 
$\xi_R$ directly.  

In order to calculate elastic scattering phase shifts, we require correlation 
functions containing two-pion interpolating operators which transform 
irreducibly according to the lattice symmetries. In particular, we require 
such correlation functions at various total momenta, and construct 
appropriate operators which transform irreducibly under the corresponding
little groups according to Ref.~\cite{Morningstar:2013bda}. While this operator construction 
procedure can be used to construct spatially-extended operators with 
gauge-covariantly displaced quark fields (which are ideal for high-lying 
resonance states), we employ only single-site hadron operators in this work. 

\begin{figure}
	\includegraphics[width=0.31\textwidth]{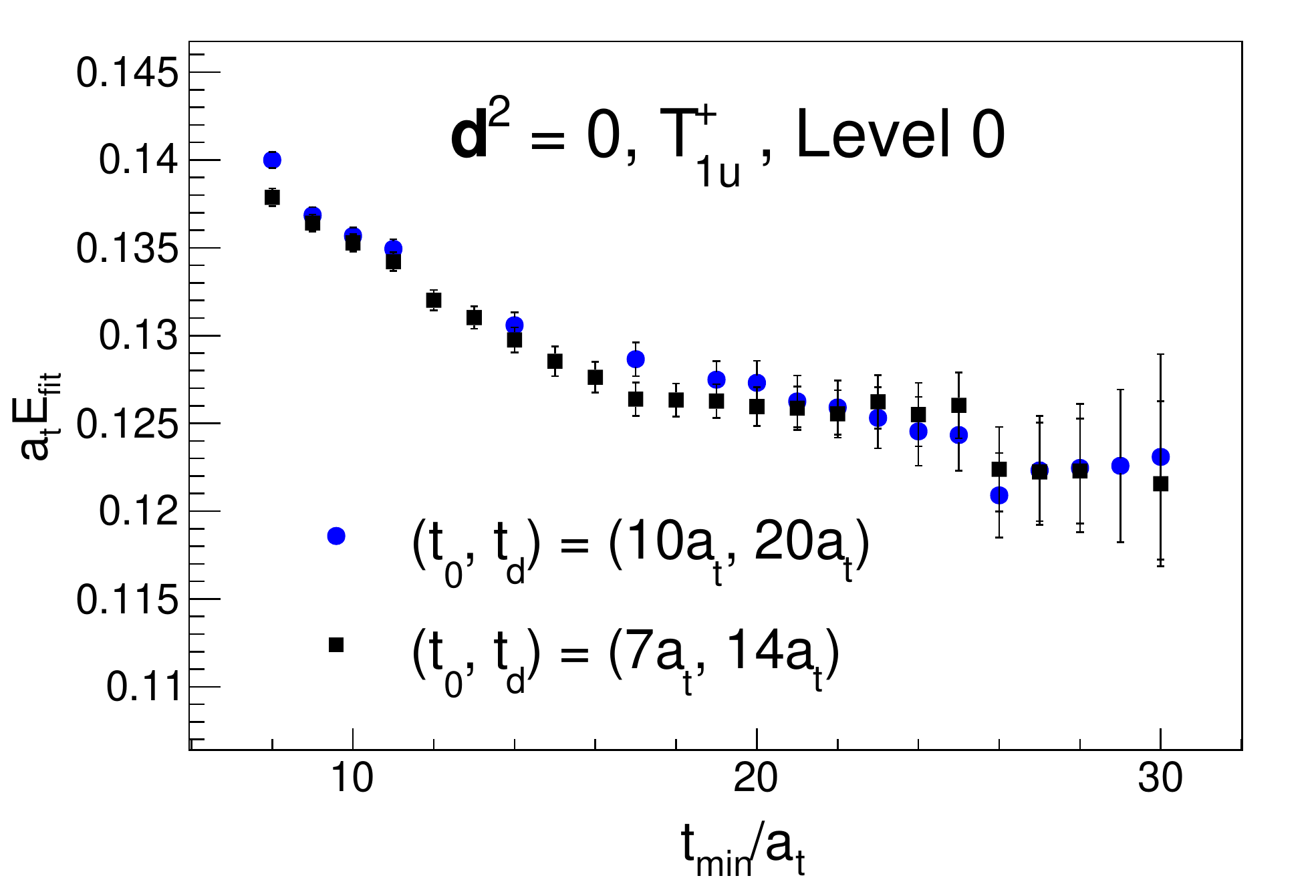}
	\includegraphics[width=0.31\textwidth]{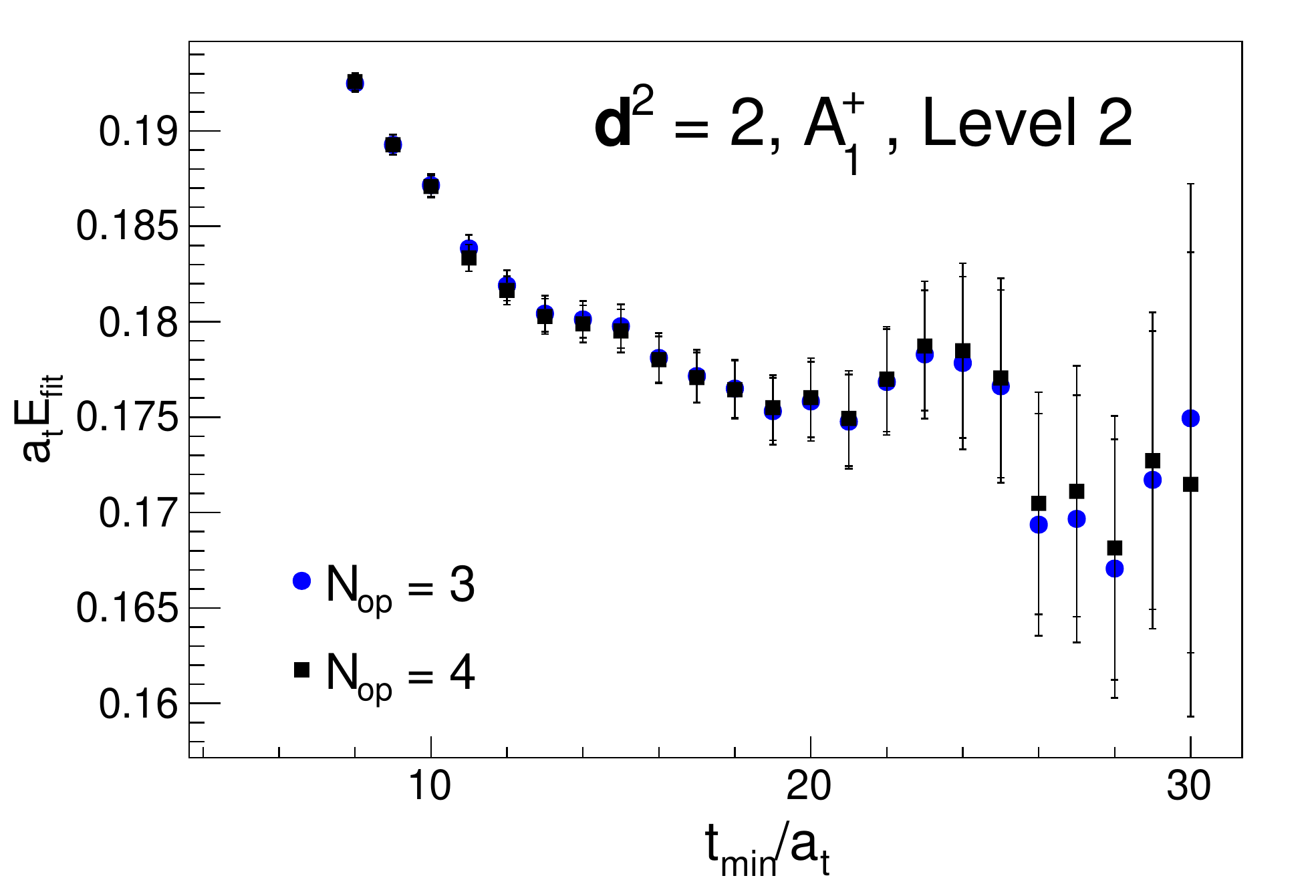}
	\includegraphics[width=0.31\textwidth]{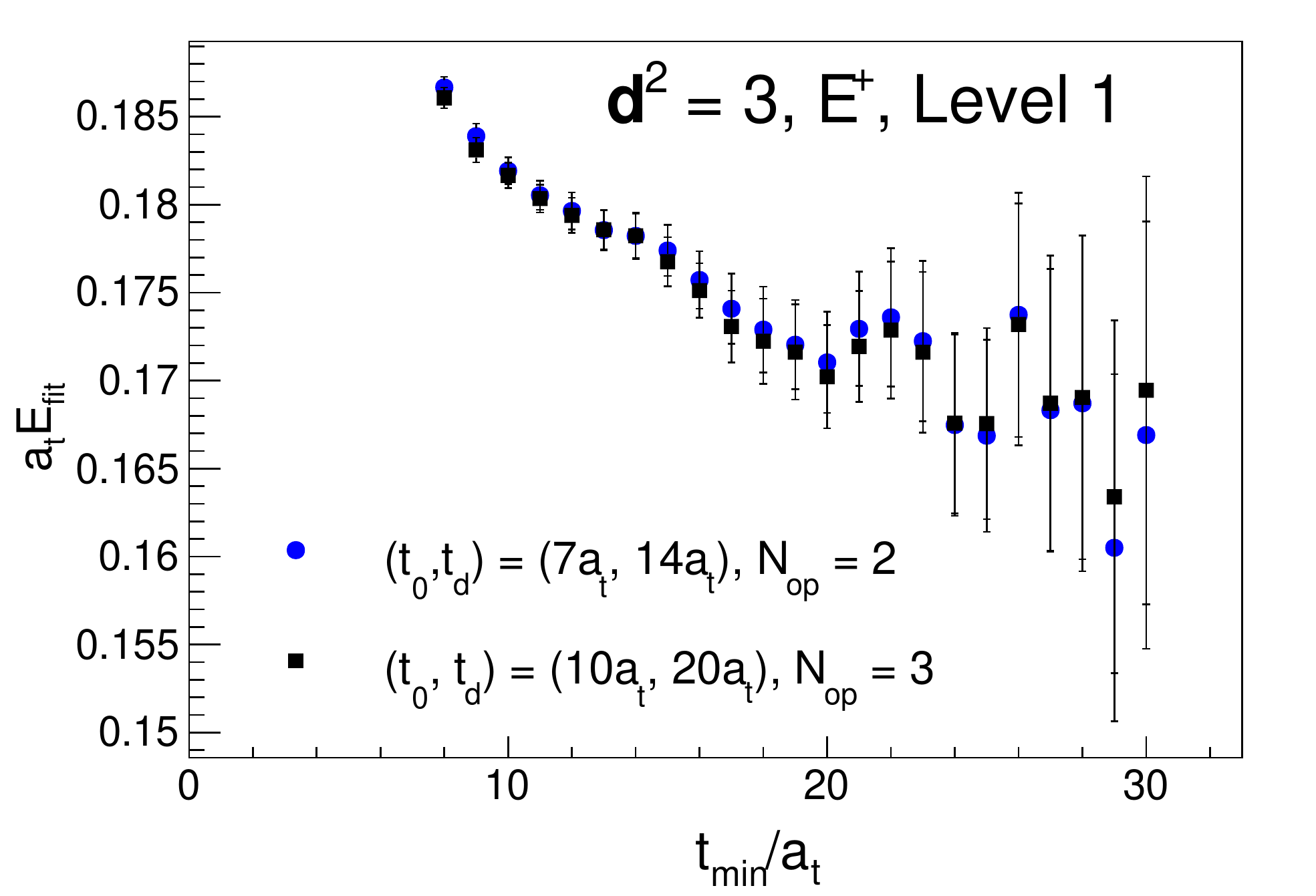}
	\caption{\label{f:tmin} Representative $t_{\rm min}$ plots from exponential 
	fits to diagonal elements of the rotated correlation matrices. Each plot 
shows results from two different choices for the GEVP parameters.}
\end{figure}

In order to extract the finite-volume energies in each irrep we 
evaluate a matrix of correlation functions 
$C_{ij}(t) = \langle O_i(t) \bar{O}_j(0) \rangle$ and 
solve the generalized eigenvalue problem (GEVP)
\begin{align}\label{e:gevp}
	C(t_d)v(t_0,t_d) = \lambda(t_0,t_d)C(t_0)v(t_0,t_d)  
\end{align}
for a single $(t_0,t_d)$. The eigenvectors $\{v_{n}(t_0,t_d)\}$ from this 
diagonalization define `optimal' interpolating operators~\cite{Michael:1982gb}
whose correlation matrix 
$\hat{C}_{ij}(t) = \langle \mathcal{O}_i (t) \bar{\mathcal{O}}_j(0) \rangle = \left( v_{i}(t_0,t_d), C(t)v_j(t_0,t_d)\right)$ is mostly 
diagonal. In order to get a preliminary idea of the spectrum, we 
perform single exponential fits to the diagonal elements of this 
rotated correlation matrix to extract the energies, taking care to vary 
$(t_0,t_d)$ to ensure that any systematic error associated 
with the off-diagonal elements of the rotated correlation matrix is smaller 
than the statistical one. We find that generally the variation of the GEVP 
parameters has little effect on the extracted energies. The fitting range 
$\left[t_{\rm min}, t_{\rm max}\right]$ is 
chosen by fixing $t_{\rm max} = 38a_t$ and varying $t_{\rm min}$ until a suitable 
$\chi^2$ is achieved and a plateau is evident. Examples of such $t_{\rm min}$ plots 
are shown in Fig.~\ref{f:tmin} which illustrate both the quality of the 
plateaux and insensitivity to the GEVP parameters.  

In addition to finite-volume energies, we can use GEVP eigenvectors to 
estimate the overlaps $Z_{in} = |\langle 0 | \hat{{O}}_i | n \rangle|^2$
between our operators and the finite-volume Hamiltonian eigenstates. 
\begin{figure}
	\includegraphics[width=\textwidth]{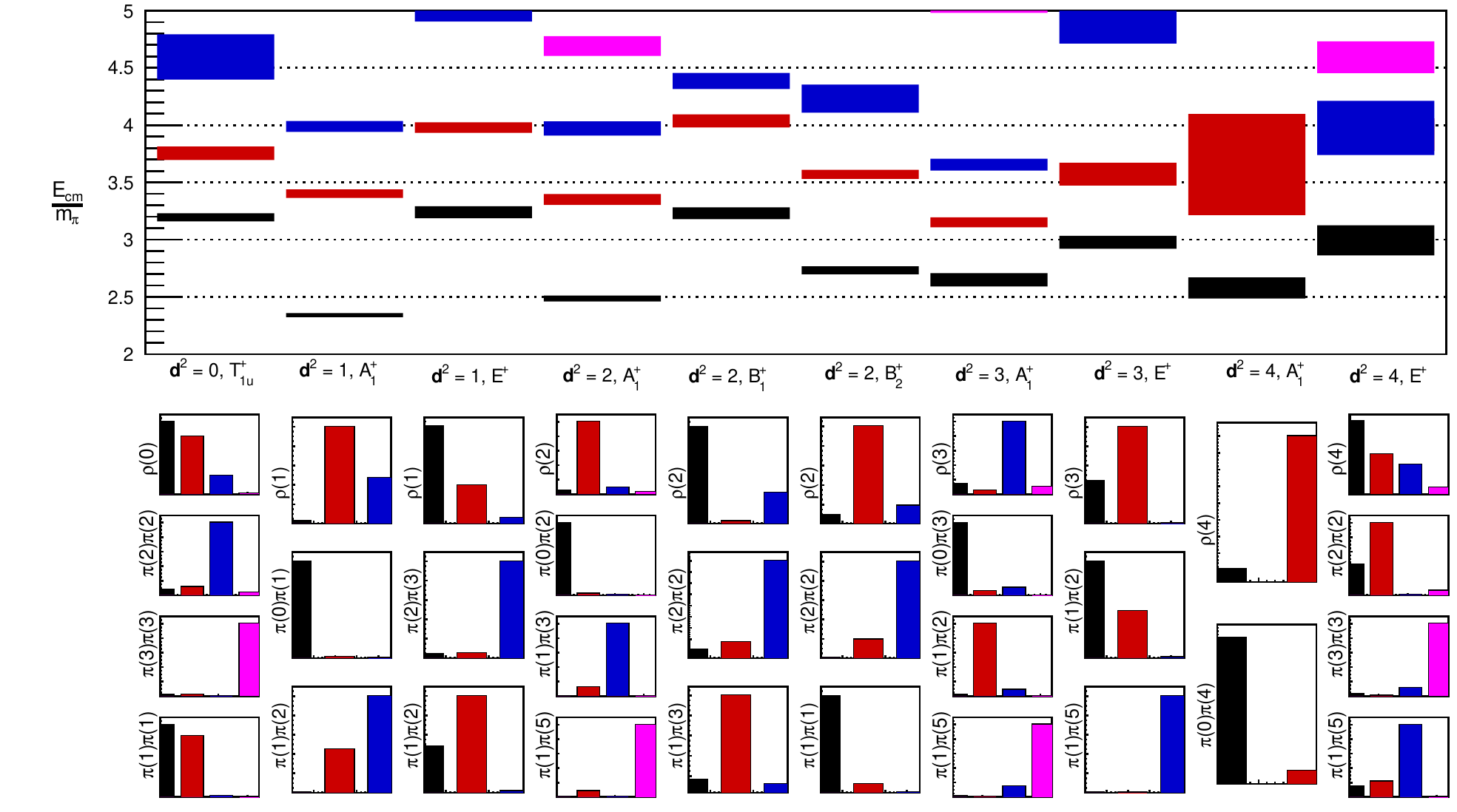}
	\caption{\label{f:i1box}$I=1$ center-of-mass energies (upper panel) for 
		each irrep together with the overlaps of each interpolator. Each column 
	corresponds to an irrep, and the bar graphs below show relative 
overlaps of each interpolating operator onto the (color-coded) finite-volume 
Hamiltonian eigenstates.}
\end{figure}
We estimate these overlaps by constructing the ratio   
\begin{align}
Z_{in}(t) = \left|\frac{\sum_{j} C_{ij}(t)v_{nj}(t_0,t_d)}{\mathrm{e}^{-\frac{E_n}{2}t}\sqrt{\hat{C}_{nn}(t)}}\right|^2, 
\end{align}
where $E_n$ is the fitted energy, 
and taking $t=20a_t$.

The $I=1$ finite-volume spectra in irreps where an infinite-volume 
$J^{PG}=1^{-+}$ state appears are shown in Fig.~\ref{f:i1box}. There we show 
center-of-mass energies $E_{\mathrm{cm}} = \sqrt{E^2 - \vec{P}^2}$, with $\vec{P} = \frac{2\pi}{L}\vec{d}$, in units of 
$m_{\pi}$ as well as overlaps onto our interpolators. As the $\rho$ meson 
is present in these irreps, we employ local single-meson operators (denoted 
$\rho(\vec{d}^2)$) as well as two-pion operators, which are denoted 
$\pi(\vec{d}_1^2)\pi(\vec{d}_2^2)$ where 
$\vec{d}^2 = (\vec{d}_1 + \vec{d}_2)^2$.  Correlation functions with equivalent
total momenta are averaged for each $\vec{d}^2$. 

A clear picture emerges from these energies and overlaps. It is only 
two-pion interpolating operators which have significant overlap with 
states far below $\frac{E_{\mathrm{cm}}}{m_{\pi}} \approx 3.2$, while single-$\rho$ interpolators have overlap for states with energies at or above this value.
Due to $G$-parity, the lowest-lying inelastic threshold is at 
$E_{\mathrm{cm}} = 4m_{\pi}$. While we are 
able to precisely extract energies near and above this threshold, they 
cannot be used to extract infinite-volume scattering information.  
Although extensions of the L\"{u}scher formula above three-hadron thresholds 
has been developed~\cite{Polejaeva:2012ut,Hansen:2014eka,Hansen:2015zga,
	Meissner:2014dea}, a rigorous treatment of four-hadron thresholds is still
	lacking. 

	In order to obtain elastic scattering phase shifts, we first define the 
	kinematic quantities   
	\begin{align}
		E_{\mathrm{cm}} = \sqrt{E^2 - \vec{P}^2}, \quad\quad \gamma = \frac{E}{E_{\mathrm{cm}}}, \quad\quad \vec{q}_{\mathrm{cm}}^2 = \frac{1}{4} E_{\mathrm{cm}}^2 - m_{\pi}^2, \quad\quad u^2 = \frac{L^2 \vec{q}_{\mathrm{cm}}^2}{(2 \pi)^2},
    \label{eqn:kinem}
  \end{align}
	where $E$ is the fitted two-hadron energy. Generally, the relation between 
	these quantities and the infinite-volume scattering matrix takes the form 
	$\mathrm{det}\; \{ 1 + F^{(\vec{d}, \gamma)}(u^2) [ S(E_{\mathrm{cm}}) - 1] \} = 0$, where $F$ is a known kinematic function and $S$ is the 
	infinite-volume scattering matrix. This relation holds up to corrections 
	which are exponential in the volume and the determinant is taken over the 
	usual $(\ell,m)$ indices of partial waves, in which $F$ is non-diagonal.

   For this work we ignore higher partial waves. Precisely, this means 
	 neglecting $\ell \ge 3$ partial waves in $I=1$ $p$-wave scattering 
	  and $\ell \ge 2$ partial waves in $I=2$ $s$-wave scattering. After applying 
    this approximation and block-diagonalizing in lattice irreps, the above 
		relation takes the particularly simple form 
		\begin{align}\label{e:quant}
			q_{\mathrm{cm}}^{2\ell +1}\mathrm{cot}\, \delta_{\ell} = -q_{\mathrm{cm}}^{2\ell +1} 
				\phi^{(\vec{d},\gamma,\Lambda)}(u^2) 
		\end{align}
	for each finite-volume irrep $\Lambda$. The functions 
	$	\phi^{(\vec{d},\gamma,\Lambda)}(u^2)$ involve 
	Rummukainen-Gottlieb-L\"{u}scher shifted zeta functions 
	($Z^{(\vec{d}, \gamma)}_{lm}(u^2)$) 
	and are given in 
	e.g. Ref.~\cite{Gockeler:2012yj} for $I=1$, $\ell=1$. For $I=2$, 
	Eq.~\ref{e:quant} takes the even simpler form 
	\begin{align}
		{q}_{\mathrm{cm}}\mathrm{cot}\, \delta_{0}(q_{\mathrm{cm}}^2) = \frac{2}{\gamma L\sqrt{\pi}}Z^{(\vec{d},\gamma)}_{00}(u^2)
	\end{align}
	for both the $A_{1g}^{+}$ and $A_{1}^{+}$ irreps used in this work. 
	It should be noted that near the two-pion threshold 
	$q_{\mathrm{cm}}^{2\ell +1}\mathrm{cot}\, \delta_{\ell}$ is analytic and 
	thus Eq.~\ref{e:quant} is valid for also for negative $q_{\mathrm{cm}}^2$. 
	In order to efficiently evaluate these zeta functions we employ a method 
	described in 
	Ref.~\cite{Fahy:2014jxa} which agrees with our implementation of 
	Appendix A in Ref.~\cite{Gockeler:2012yj}. 

While the single-exponential fits discussed above provide preliminary spectra, 
the center-of-mass momentum and thus the scattering phase shifts are very 
sensitive to these energies. For two-hadron dominated levels, the energies 
will be very close to their non-interacting values. 
Therefore, for these levels 
it is beneficial to perform fits to the ratio of two and 
single-hadron correlators (as in Ref.~\cite{Helmes:2015gla} but here 
generalized to arbitrary momenta) 
\begin{align}\label{e:ratio}
	R(t) = \frac{\langle \mathcal{O}_{\vec{d}^2_1,\vec{d}^2_2} (t)
	\bar{\mathcal{O}}_{\vec{d}_1^2,\vec{d}^2_2}(0) \rangle}{
		\langle O_{\vec{d}_1}(t) 
		\bar{O}_{\vec{d}_1}(0)\rangle
		\langle O_{\vec{d}_2}(t) 
	\bar{O}_{\vec{d}_2}(0)\rangle},
\end{align}
where $\hat{\mathcal{O}}_{\vec{d}^2_1,\vec{d}^2_2}$ is a optimized operator for an eigenstate 
dominated by   
individual pions with momenta of magnitude $\vec{d}^2_1$ and $\vec{d}_2^2$, 
respectively. The $O_{\vec{d}}$ are single-pion operators with momentum 
$\vec{d}$. Single exponential fits to these ratios directly yield 
this energy difference, and typically have less excited state 
contamination than ordinary single exponential fits. However, the 
excited state contamination in such fits may not 
be monotonically decreasing, which can complicate the identification of 
the plateau region in some cases. 
Another method to better resolve differences between 
the two- and single-hadron energies is to perform 
simultaneous fits to the two hadron correlators included in the ratio of 
Eq.~\ref{e:ratio}.  
While both methods yield consistent results, we quote values from   
the simultaneous fits, as they result in more conservative statistical errors. 

\begin{figure}
	\includegraphics[width=0.49\textwidth]{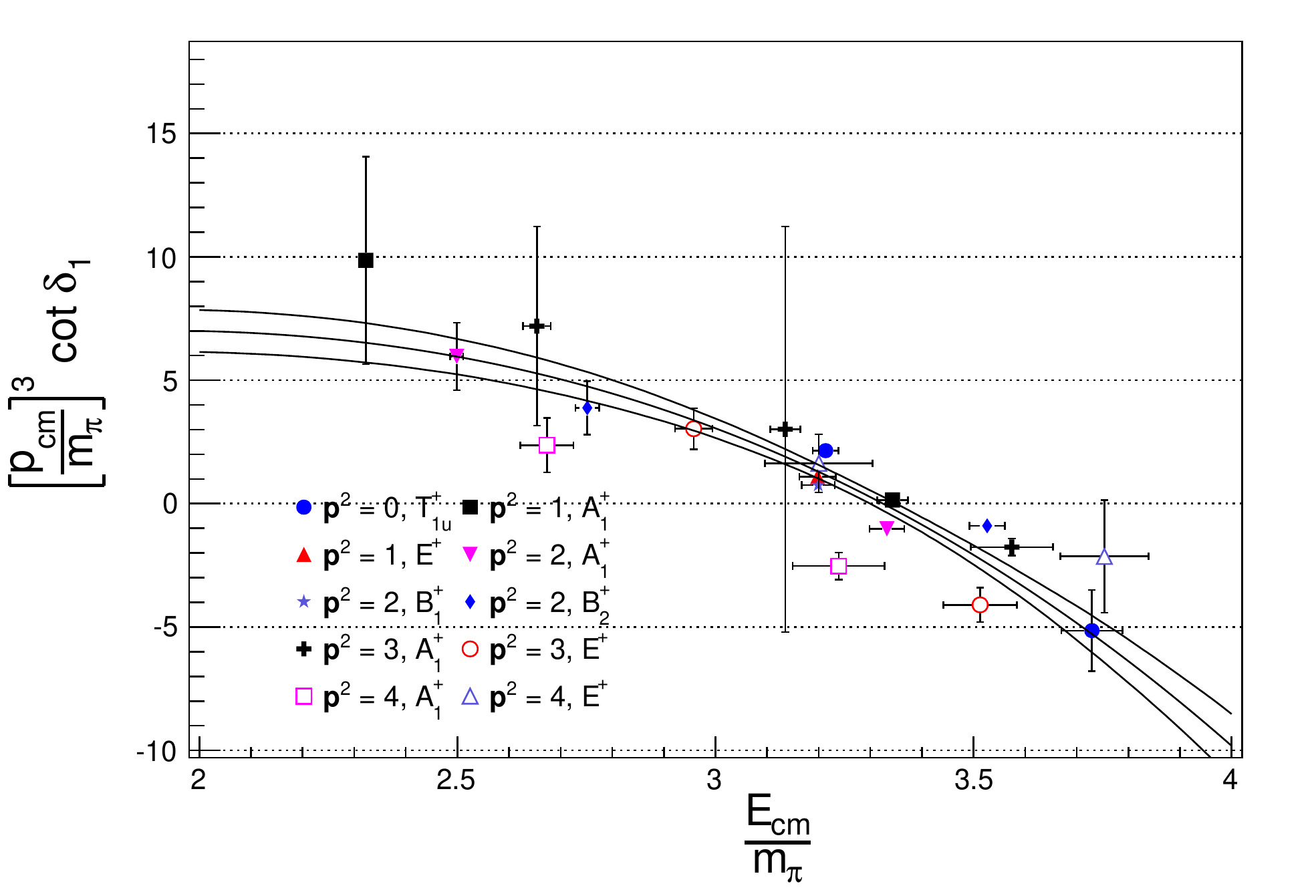}
	\includegraphics[width=0.49\textwidth]{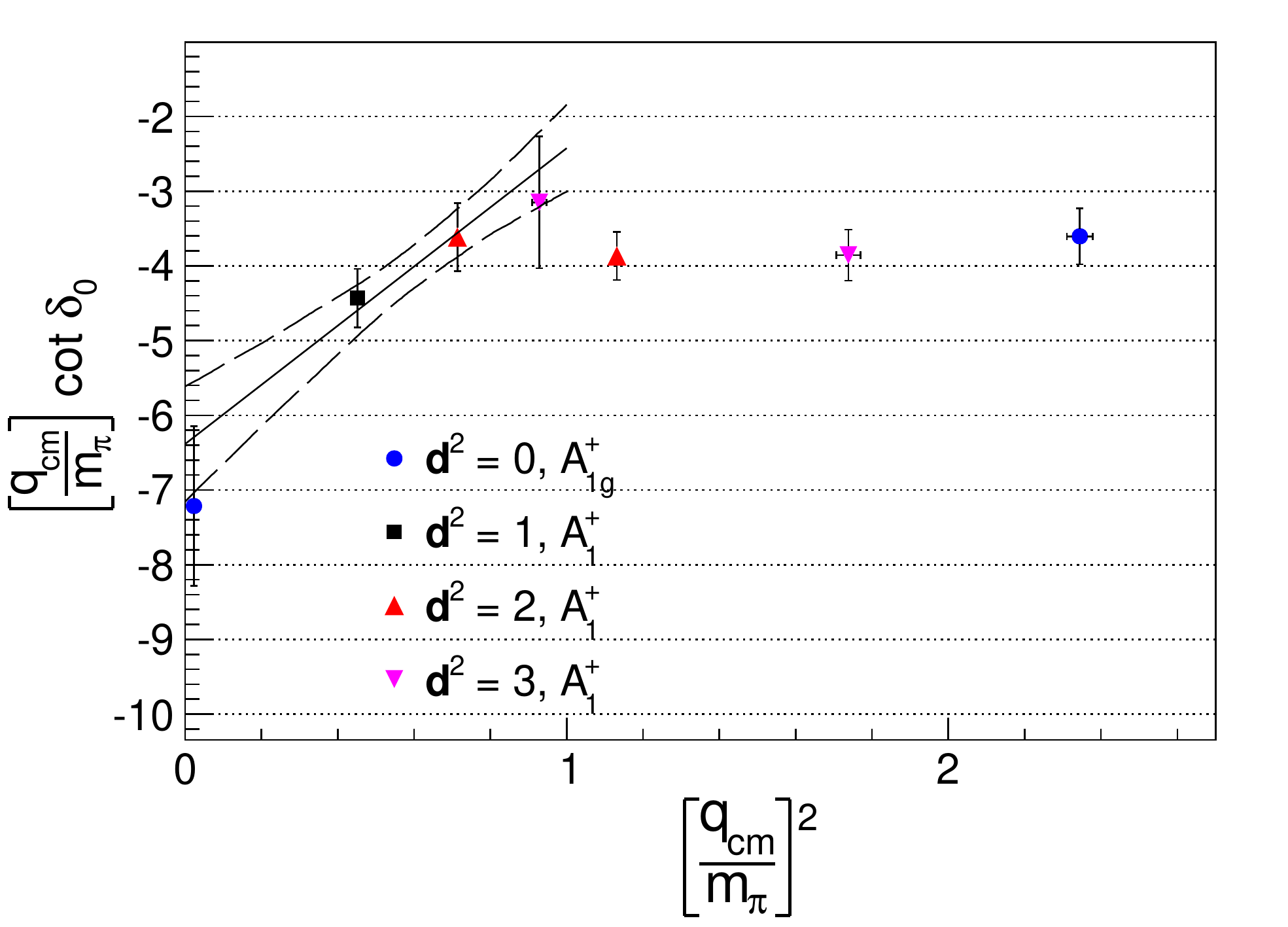}
	\caption{\label{f:i1scat} The $I=1$ $p$-wave scattering phase shift (left) 
		and $I=2$ $s$-wave phase shift (right) together with fits to 
		$q_{\mathrm{cm}}^{2\ell+1} \cot \delta_{\ell}$ described in the text.}
\end{figure}
The resultant scattering phase shifts are shown in Fig.~\ref{f:i1scat} for 
both $I=1$ and $I=2$, together with fits to 
$q_{\mathrm{cm}}^{2\ell+1} \cot \delta_{\ell}$. Care must be taken to treat the
correlation between $x-$ and $y$-errors as well as among the data points 
in these fits. Upon every call to the correlated $\chi^2$ function, we estimate
the necessary covariance matrix using the $N_b=800$ bootstrap samples of these 
secondary observables. The minimization is performed on each bootstrap sample, 
where bootstrap replica of the means are used together with the covariance 
matrix to construct the correlated $\chi^2$. For $I=1$ $p$-wave scattering, 
we fit to a relativistic Breit-Wigner form 
\begin{align}
    \left( \frac{q_{cm}}{m_\pi} \right)^3 \cot \delta_1 = \left( \frac{m_\rho^2}{m_\pi^2} - \frac{E_{cm}^2}{m_\pi^2} \right) \frac{6 \pi E_{cm}}{g_{\rho \pi \pi}^2 m_\pi}
    \label{eqn:bwigner}
	\end{align}
and obtain
\begin{align}
	g_{\rho\pi\pi} = 6.16(36),\quad   \frac{m_{\rho}}{m_{\pi}} = 3.324(24), \quad
	\frac{\chi^2}{d.o.f.} = 1.43
\end{align}
where the coupling is in good agreement with the experimental value. For $I=2$ 
$s$-wave scattering, we employ the NLO effective range expansion to the 
points at low momenta $q_{\mathrm{cm}} \le m_{\pi}$ 
\begin{align}\label{e:er}
	\left(\frac{q_{\mathrm{cm}}}{m_{\pi}}\right)\cot \delta_0 = \frac{1}{m_{\pi}a_{0}} + 
	\frac{1}{2}(m_{\pi}r)\left( \frac{q_{\mathrm{cm}}}{m_{\pi}}\right)^2  
\end{align}
and obtain 
\begin{align}\label{e:i2fit}
	m_{\pi} a^{I=2}_{0} = -0.157(19) ,\qquad m_{\pi} r = 7.9(2.4), \qquad \chi^2/d.o.f. = 0.61. 
\end{align}

The JLab group~\cite{Wilson:2015dqa} has recently calculated the 
$I=1$ $p$-wave phase shift on this same ensemble using the distillation method 
of Ref.~\cite{Peardon:2009gh}. While their results are more precise than those
quoted here, significantly more Dirac matrix inversions are required. Using 
the notation of Ref.~\cite{Morningstar:2011ka}, we employ 
(TF, SF, LI8) dilution for fixed quark lines and (TI16, SF, LI8)
for relative ones. Our use of five fixed and two relative lines (with eight 
source times) results in $N_D = 2304$ Dirac matrix inversions for each gauge 
configuration, while Ref.~\cite{Wilson:2015dqa} requires $N_D = 393216$ for 
the distillation method.

\section{$I=1$ phase shifts and the timelike pion form factor on a 
CLS ensemble}\label{s:ff}

	Motivated by the efficacy of the stochastic LapH method in the large-volume 
	calculation described in Sec.~\ref{s:aniso}, we now employ it to isotropic
	$N_{\mathrm{f}} = 2+1$ ensembles 
	generated within the CLS community effort~\cite{Bruno:2014jqa}. These 
	ensembles have a variety of lattice spacings, physical volumes, and 
	pion masses enabling an assessment of these systematic errors. 
	As a first step in this direction, here we use a single ensemble (the N200) 
	which is described in Tab.~\ref{t:ens}. 
	\begin{table}
		\centering
	\begin{tabular}{|c|c|c|c|c|}
		\hline
		$(L/a)^3\times(T/a)$ & $m_{\pi} (\mathrm{MeV})$ & $a (\mathrm{fm})$ & 
		 $m_{\pi}L$ & $N_{\mathrm{cfg}}$ \\
		\hline
		$48^3 \times 128$ & $280$ & $0.065$ & $4.3$ & $1710$ \\ 
		\hline
	\end{tabular}
	\caption{\label{t:ens}Details for the isotropic $N_{\mathrm{f}} = 2+1$ gauge 
		configurations used here. For a complete discussion of the 
		ensemble generation and preliminary 
		scale setting, see Ref.~\cite{Bruno:2014jqa}.}
\end{table}

Rather than the periodic (anti-periodic) temporal boundary conditions for 
bosons (fermions) which are typically employed in the lattice QCD simulations, 
these ensembles use `open' temporal boundary conditions~\cite{Luscher:2011kk} 
which improve the scaling of the exponential 
autocorrelation time $\tau_{\mathrm{exp}}$ as the continuum limit is 
approached. This means source and sink times must be chosen 
carefully in order to avoid boundary effects. Based on the 
observables considered in Refs.~\cite{Bruno:2014jqa,Bruno:2014lra} and on the 
behavior of the LapH eigenvalues, we use $t_0 = T/4 = 32a$. 

In the gauge-covariant 3-D Laplace operator used to define the LapH subspace we 
stout smear~\cite{Morningstar:2003gk} the gauge links with parameters 
$(\rho, n_{\rho}) = (0.1, 36)$. We choose a LapH cutoff 
$\sigma_s \approx 1\mathrm{GeV}$ which 
results in a cutoff eigenvalue of $(a\sigma_s)^2 \approx 0.11$ and $N_v = 192$.
The physical volume is somewhat smaller on this lattice compared to the 
anisotropic ensemble of Sec.~\ref{s:aniso}, so a reduced $N_v$ is 
expected. We employ the same (TF, SF, LI8) dilution scheme for fixed lines, 
but for relative lines use (TI8, SF, LI8) as the temporal lattice spacing 
is larger. We use four independent fixed quark lines, a single relative line, 
and one source time $t_0 = 32a$. The $N_D = 384$ Dirac matrix inversions per 
configuration are performed using the DFL-SAP-GCR solver~\cite{Luscher:2007se} 
implemented in 
\texttt{openQCD}\footnote{http://luscher.web.cern.ch/luscher/openQCD/}, which 
we have integrated into the stochastic LapH codebase. The results from these 
inversions are projected onto the LapH subspace and stored for later use in 
other calculations, such as Ref.~\cite{koch}. Finally, these ensembles employ 
RHMC and twisted mass re-weighting, and we multiply all primary observables
by the corresponding re-weighting factors. 

The renormalization and $\mathrm{O}(a)$ improvement of composite operators 
is simplified on an isotropic lattice. Furthermore, the regularization employed 
by these CLS ensembles is well studied and many renormalization and 
improvement coefficients have been previously determined. 
Therefore, in addition to applying the methods of Sec.~\ref{s:aniso} to 
calculate the $I=1$ $p$-wave scattering phase shift, we also calculate a  
matrix element of the vector current with two-pion states. 
The simplest such matrix element of phenomenological relevance is the timelike 
pion form factor. As before we are restricted to the elastic region which, for 
this ensemble is $2m_{\pi} \le E_{\mathrm{cm}} \le 2m_{K}$. In this region the 
timelike pion form factor
$|F_{\pi}(E_{\mathrm{cm}})|^2$ can be defined as~\cite{Jegerlehner:2009ry} 
\begin{align}\label{e:ffdef}
	R(s) \equiv \frac{\sigma(e^{+}e^{-} \rightarrow hadrons)}{4\pi\alpha(s)^2/(3s)} = 
	\frac{1}{4}\left( 1 - \frac{4m_{\pi}^2}{s}\right)^{\frac{3}{2}}|F_{\pi}(\sqrt{s})|^2,  
\end{align}
where the denominator in $R(s)$ is the tree-level cross section 
$\sigma(e^{+}e^{-} \rightarrow \mu^{+}\mu^{-})$ for 
$s = E_{\mathrm{cm}}^2 \gg m_{\mu}^2$. Effectively, this form factor  
describes QCD corrections to the coupling of a (virtual) 
photon to two pions. It is phenomenologically relevant because of (e.g.) 
its relation to the low-energy contribution to the hadronic vacuum 
polarization (HVP) 
$\Pi(Q^2)$. For spacelike four-momentum transfer $Q^2$, the once-subtracted 
dispersion relation 
\begin{align}\label{e:hvp}
	\Pi(0) - \Pi(Q^2) = Q^2 \int_{0}^{\infty} ds \, \frac{\rho(s)}{s(s+Q^2)}, \quad \rho(s) = \frac{R(s)}{12\pi^2}
\end{align}
expresses the HVP in terms of $R(s)$. Typically, this dispersion relation 
is not used in lattice QCD calculations of the HVP which is instead calculated 
directly from current-current correlation functions. 
Furthermore, the relation between $R(s)$ and $|F_{\pi}(E_{\mathrm{cm}})|^2$ 
given in Eq.~\ref{e:ffdef} is valid only in the elastic region, while the 
integral in Eq.~\ref{e:hvp} is unbounded from above. However, direct 
lattice calculations of the HVP require fully disconnected Wick contractions (which are typically ignored), and suffer from large statistical errors 
and finite-volume effects in the low-$Q^2$ region~\cite{Aubin:2015aya}. 
Therefore, a 
`hybrid' determination which combines lattice data from both $\Pi(Q^2)$ and 
$|F_{\pi}(E_{\mathrm{cm}})|^2$ may be the best approach\footnote{We thank Harvey Meyer for clarifying this point.}. 

In analogy with earlier work by Lellouch and Luscher~\cite{Lellouch:2000pv}, a 
relation between the infinite-volume $|F_{\pi}(E_{\mathrm{cm}})|^2$ and finite-volume matrix elements (up to exponential finite-volume corrections) was 
derived by Meyer in Ref.~\cite{Meyer:2011um}  
\begin{align}
	\left | F_\pi(E_{\mathrm{cm}}) \right |^2 =& 
	g^{(\vec{d}, \Lambda)}(\gamma) \left( 
		u \frac{\mathrm{d} \phi^{(\vec{d}, \Lambda)}(u^2)}{\mathrm{d} u} + 
	q_{\mathrm{cm}} \frac{\partial \delta_1(q_{\mathrm{cm}})}{\partial q_{\mathrm{cm}}} 
\right) \frac{3 \pi E_{\mathrm{cm}}^2}{2 q_{\mathrm{cm}}^5} 
| A^{(\vec{d}, \Lambda)} |^2, \label{e:meyer} \\\nonumber
		& g^{(\vec{d}, \Lambda)}(\gamma) = \begin{cases} 1 / \gamma &\mbox{if } \Lambda = A_1 \\ \gamma &\mbox{else} \end{cases}, 
  \end{align}
	where $\phi^{(\vec{d}, \Lambda)}(u^2)$ is given in Eq.~\ref{e:quant} and 
\begin{align}
	A^{(\vec{d}, \Lambda)} = \braket{0|V^{(\vec{d}, \Lambda)}|\vec{d} \, \Lambda \, n}
    \label{eqn:currmatel}
  \end{align} 
is a finite-volume matrix element involving an $I=1$ two-pion state with 
total momentum $\vec{d}$ in irrep $\Lambda$ below inelastic threshold. While 
this relation was derived only for total zero momentum in 
Ref.~\cite{Meyer:2011um}, it may be straight-forwardly extended to non-zero 
total momentum using the argumentation of Ref.~\cite{Briceno:2015csa}.
A proof-of-principle application of this relation was performed recently in 
Ref.~\cite{Feng:2014gba} while a similar matrix element is calculated in 
Ref.~\cite{Briceno:2015dca}. 

Since 
we work in the isospin limit, the electromagnetic current 
$\hat{J}^{\mathrm{em}}_i = \frac{2}{3}\bar{u}\gamma_i u - \frac{1}{3} \bar{d} \gamma_i d - \dots$ is replaced by the isospin current 
\begin{align}
	V_i^{a} = \bar{\psi}\gamma_i \frac{\tau^{a}}{2} \psi, \quad \psi = \left( {u \atop d} \right),  
\end{align}
where $\tau^{a}$ is an $SU(2)$ generator. Furthermore, we project this current 
onto finite-volume irreps by defining $V^{(\vec{d}, \Lambda)} = 
b_i^{(\vec{d}, \Lambda)} V_i$. The vectors $\vec{b}^{(\vec{d}, \Lambda)}$ are 
given in Tab.~\ref{t:currirr} for all irreps used in this work. 
  \begin{table}
    \centering
    \begin{tabular}{c c c}
      \hline
			Reference momentum $\vec{d}$ & Irrep $\Lambda$ & $\vec{b}$ \\
      \hline \hline
      [000] & $T_{1u}$ & (1,0,0) \\
      \hline
      [00n] & $A_{1}$ & (0,0,1) \\
            & $E$ & (0,1,0) \\
      \hline
      [0nn] & $A_{1}$ & $\frac{1}{\sqrt{2}}$ (0,1,1) \\
            & $B_1$ & (1,0,0) \\
            & $B_{2}$ & $\frac{1}{\sqrt{2}}$ (0,-1,1) \\
      \hline
      [nnn] & $A_{1}$ & $\frac{1}{\sqrt3}$ (1,1,1) \\
            & $E$ & $\frac{1}{\sqrt{2}}$ (1,-1,0) \\
    \end{tabular}
		\caption{Linear combinations of components of the vector current such that $V^{(\vec{d}, \Lambda)} =  b^{(\vec{d}, \Lambda)}_i V_i$ transforms irreducibly 
		according to the irrep $\Lambda$.}
    \label{t:currirr}
  \end{table}
  In order to obtain the derivative of the $p$-wave phase shift required in 
	Eq.~\ref{e:meyer}, we first perform the complete phase shift analysis and 
	parametrize the energy dependence of $\delta_1$ using the Breit-Wigner form 
	of Eq.~\ref{eqn:bwigner}.  
	The derivative is then evaluated on each bootstrap sample and taken 
	together with bootstrap samples of the current matrix elements of 
	Eq.~\ref{eqn:currmatel} to obtain samples of $|F_{\pi}(E_{\mathrm{cm}})|^2$.

  In order to obtain the current matrix elements in 
	Eq.~\ref{eqn:currmatel}, we must evaluate correlation functions of the form
  \begin{align}
		D_i(t) = \braket{V^{(\vec{d}, \Lambda)}(t_0 + t) \bar O^{(\vec{d}, \Lambda)}_i(t_0)},
    \label{e:ccorr}
  \end{align}
	where the $\{ O^{(\vec{d}, \Lambda)}_i \}$ are interpolators used in the 
	GEVP of Eq.~\ref{e:gevp}. In order to ensure an $O(a^2)$ approach to the 
	continuum limit, the currents $V^{(\vec{d}, \Lambda)}$ are linear 
	combinations  of the renormalized,  
	$\mathcal{O}(a)$-improved local vector current $(V_{\mathrm{R}})_\mu$ defined as~\cite{Luscher:1996sc} 
  \begin{align}
		(V_{\mathrm{R}})^{a}_{\mu} = Z_{\mathrm{V}}(1 + b_{\mathrm{V}}\,am_{\mathrm{q}})
		\{ V^{a}_{\mu} + ac_{\mathrm{V}}\tilde{\partial}_{\nu}T^{a}_{\mu\nu}\},
    \label{e:vimp}
  \end{align}
  where $Z_{\mathrm{V}}$, $ b_{\mathrm{V}}$, and $ac_{\mathrm{V}}$ are 
	renormalization and improvement coefficients, $am_{\mathrm{q}}$  
	the bare quark mass in lattice units, $T^{a}_{\mu\nu}= i\bar{\psi} \sigma_{\mu\nu}\frac{\tau^{a}}{2}\psi$, and $\tilde{\partial}_{\nu}$ the symmetrized lattice 
	derivative. A preliminary determination of the renormalization coefficient
	$Z_{\mathrm{V}}$ for the CLS lattice regularization
	has been provided in Ref.~\cite{mattia}, while 
	$b_{\mathrm{V}}$, and $ac_{\mathrm{V}}$ are calculated to 1-loop in 
	Ref.~\cite{Aoki:1998qd}. For this preliminary work, we subsitute 
	the unrenormalized PCAC mass $am_{\mathrm{PCAC}}$ for $am_{\mathrm{q}}$ 
	which holds at tree level, so that $O(a)$ improvement is formally 
	implemented only at this order. However, this effect is suppressed by 
	the small quark mass.  

  In the stochastic LapH framework interpolators are built from 
	quark fields projected onto the LapH subspace, while quark fields 
	appearing in the current are local. However, this can be easily accommodated
	by forming the meson functions in Eq.~32 of Ref.~\cite{Morningstar:2011ka} 
	with quark sinks which are not projected into the LapH subspace. The 
	unprojected `current' functions are otherwise completely analogous to the 
	meson functions in the construction of correlators but are calculated 
	immediately after the Dirac matrix inversions but before the sinks are 
	projected and written out to disk. In practice, correlation functions
	required for the matrix elements  
	\begin{align}\label{e:me}
		A_{n}^{(0)} =  \langle 0|V^{(\vec{d},\Lambda)}|\vec{d}\Lambda n\rangle, \qquad 
		A_{n}^{(1)} =  \langle 0|b_{i}^{(\vec{d},\Lambda)}\, \tilde{\partial}_{\nu}T^{a}_{i\nu} |\vec{d}\Lambda n \rangle
	\end{align}
	are calculated and analyzed separately.

	We turn now to extraction of the matrix elements given Eq.~\ref{e:me}. 
	In analogy with our procedure for the energies, we measure the 
	correlation functions of Eq.~\ref{e:ccorr} and with the GEVP eigenvectors 
	form their 
	`optimized' counterparts 
		$\hat{D}_i(t) = ( D(t), v_i(t_0, t_d))$, 
	where the inner product is taken over the GEVP indices. Up to GEVP 
	corrections (treated as a systematic error as described in the previous 
	section) these optimized 
	correlation functions have the large-time behavior
	\begin{align}
		\lim_{t\rightarrow\infty} \hat{D}_i(t) = \braket{0|V^{(\vec d, \Lambda)}|\vec{d}\Lambda i} \braket{\vec{d} \Lambda i| \bar O^{(\vec d, \Lambda)}_i|0} \times \mathrm{e}^{-E^{(\vec{d},\Lambda)}_i (t-t_0)}.
    \label{eqn:optcurrcorrdec}
  \end{align}
  This suggests three different ratios which tend to the matrix elements of 
	interest:
  \begin{align}\label{eqn:ratios}
		R_1^{(i)}(t) = \frac{\hat D_i(t)}{{\hat{C}}_{ii}^{\frac12}(t) \mathrm{e}^{-\frac12 E_i (t-t_0)}}, \quad
		R_2^{(i)}(t) = \frac{\hat D_i(t) \braket{\vec{d}\Lambda n|\bar {O}^{(\vec d, \Lambda)}_i|0}}{{\hat{C}}_{ii}(t)}, \quad
		R_3^{(i)}(t) = \frac{\hat D_i(t)}{\braket{\vec{d}\Lambda n|\bar {O}^{(\vec d, \Lambda)}_i|0} \mathrm{e}^{-E_i (t-t_0)}},
  \end{align}
	where the overlaps and energies appearing in these expressions are obtained 
	from fits to the diagonal elements of the rotated correlation matrix. 
	The value of the matrix element is taken to be a plateau average of the ratio
	over a suitable region. 
	Alternatively, simultaneous fits to $\hat{C}_{ii}(t)$ and $\hat{D}_i(t)$ may
	also be used to extract the desired matrix elements. Generally we find that 
	all four of these different determinations yield results which are consistent
	within statistical errors for fit ranges in the plateau region.  

	\begin{figure}
		\includegraphics[width=0.49\textwidth]{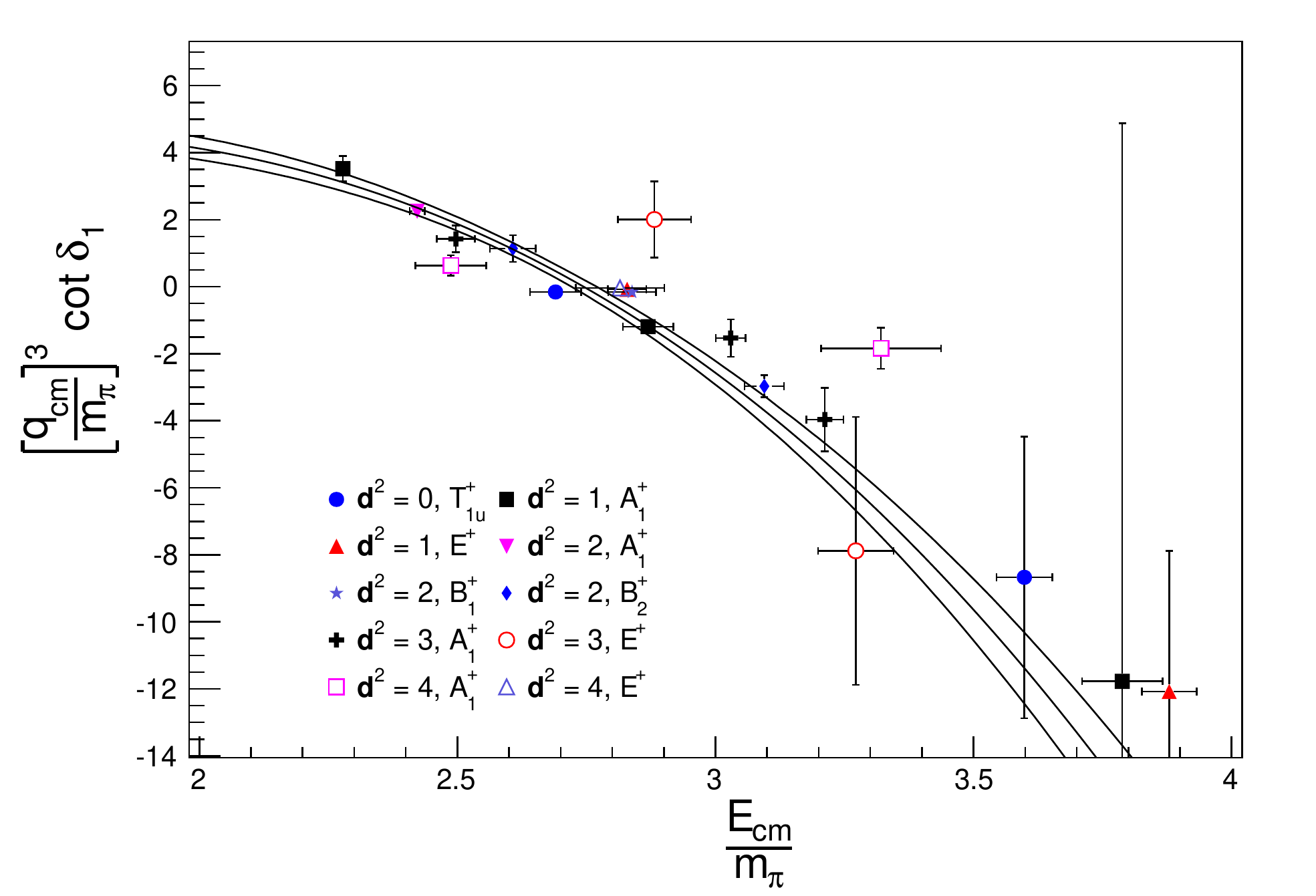}
		\includegraphics[width=0.49\textwidth]{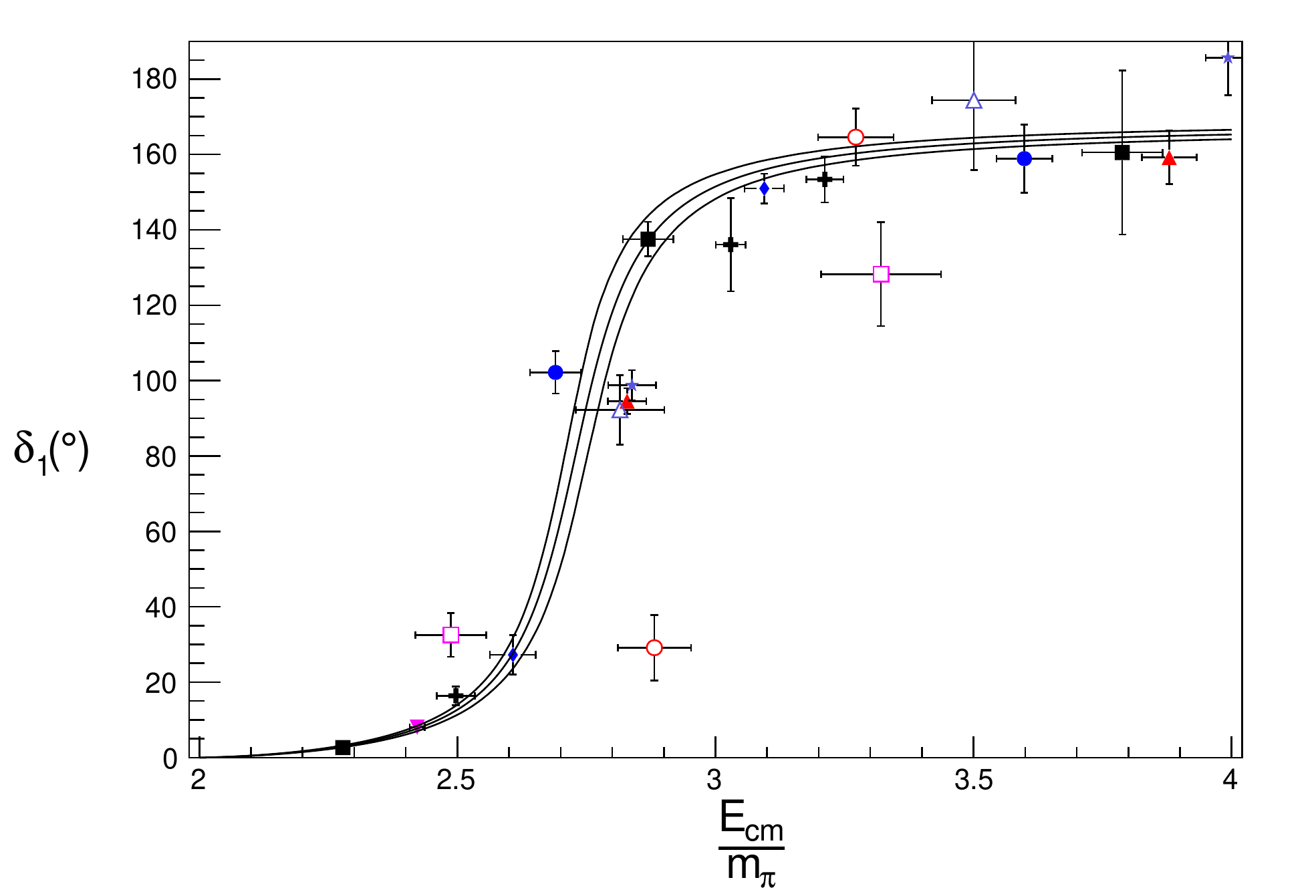}
		\caption{\label{f:phase48} The $p$-wave scattering phase shift on the 
			N200 CLS lattice. Shown are $q_{\mathrm{cm}}^3\cot{ \delta_1}$ (left) 
			and $\delta_1$ (right) together with a Breit-Wigner fit to 
			$q_{\mathrm{cm}}^3\cot{ \delta_1}$. The resultant fit parameters are 
		given in the text. The lowest inelastic threshold is at $2m_K \approx 3.2m_\pi$. Points above this threshold are shown on the graphs but not 
	included in the fit.}
	\end{figure}
\begin{figure}
		\includegraphics[width=0.49\textwidth]{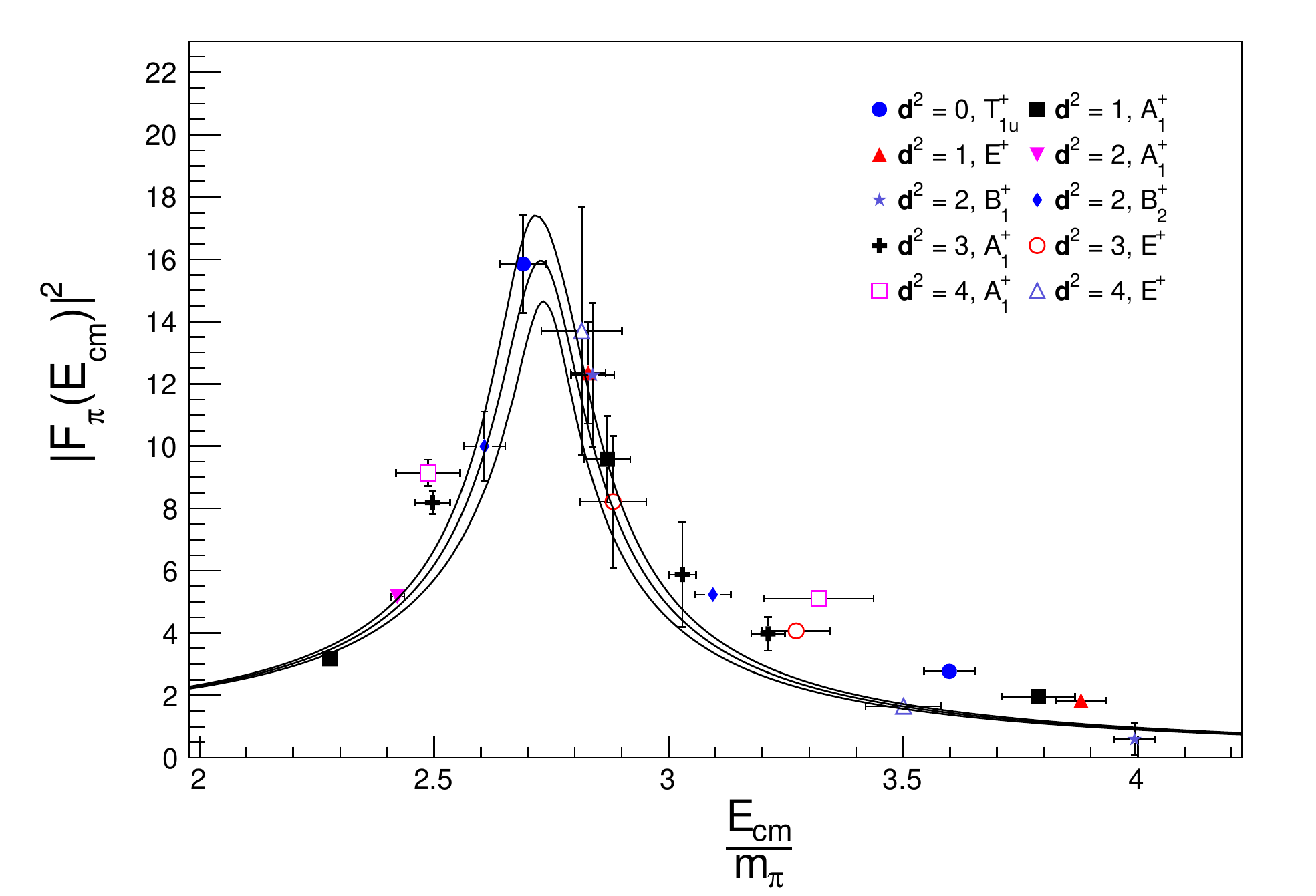}
		\includegraphics[width=0.49\textwidth]{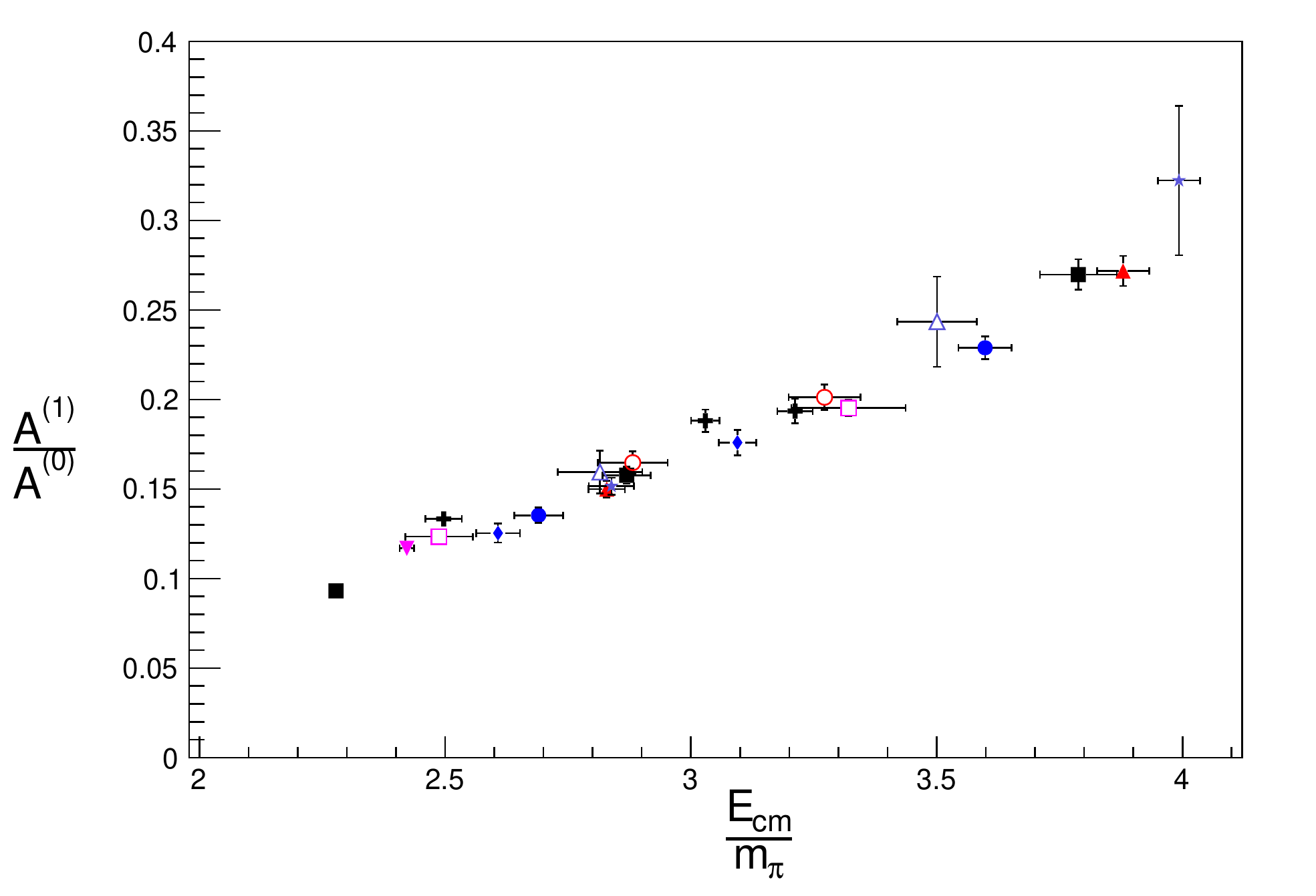}
		\caption{\label{f:ff}The timelike pion form factor together with 
		the expected Gounaris-Sakurai parametrization using the previously-calculated $m_{\rho}$ and $g_{\rho\pi\pi}$ (left). The ratio of the matrix element 
	which contributes at $O(a)$ over the leading order one is shown on the right 
for each of the form factor data points.}	
	\end{figure}

	Our results from the phase shift determination on this lattice are shown 
	in Fig.~\ref{f:phase48}. 
	As mentioned above, the lowest inelastic threshold in this channel is due to 
	two kaons at $2m_{K} \approx 3.2m_{\pi}$. Although we have a number of energy levels above this threshold, they are not included in the final analysis but 
	shown for illustration. As before, we fit 
	$q_{\mathrm{cm}}^{3}\cot \delta_1$ to the Breit-Wigner form
	of Eq.~\ref{eqn:bwigner} which gives 
	\begin{align}
	g_{\rho\pi\pi} = 5.68(24),\quad   \frac{m_{\rho}}{m_{\pi}} = 2.745(24), \quad
	\frac{\chi^2}{d.o.f.} = 1.20
	\end{align}
	which is somewhat lower than the experimental value of $g_{\rho\pi\pi}$. 
	The Breit-Wigner parametrization of the phase shift in 
	Eq.~\ref{e:meyer} together with the matrix elements $A^{(0)}$, $A^{(1)}$ and 
	 renormalization/improvement coefficients of Eq.~\ref{e:vimp} enables the 
	extraction $|F_{\pi}(E_{\mathrm{cm}})|^2$, which is shown in Fig.~\ref{f:ff}.
	Also shown in that figure is the ratio of the matrix element
	appearing in the $O(a)$ term over the leading order one. We see that this 
	$O(a)$ matrix element grows from $\sim 10\%$ of the leading one at low 
	momenta to $\sim 30\%$ at our largest momentum. Due to the small 1-loop 
	value of $ac_{\rm V}$, this 
	term therefore has no significant effect on our final results. However, 
	non-perturbative determinations of these improvement coefficients can be 
	considerably larger than the 1-loop value~\cite{Bulava:2015bxa} possibly 
	increasing the influence of this term to the few-percent level. In future work
, an alternative calculation which employs the point-split vector current will 
give an additional handle on the magnitude of these $O(a)$ effects. 

	Also shown in Fig.~\ref{f:ff} is the Gounaris-Sakurai parametrization of 
	$|F_{\pi}(E_{\mathrm{cm}})|^2$ which (using the notation of 
	Ref.~\cite{Francis:2013qna}) is  
	\begin{align}
		F_{\pi}^{GS}(\sqrt{s}) = &\frac{f_0}{q_{\rm cm}^2h(\sqrt{s}) - q_{\rho}^2 h(m_{\rho})
		+ b(q_{\rm cm}^2 - q_{\rho}^2) - \frac{q_{\rm cm}^3}{\sqrt{s}}i},
		\\\nonumber 	
		b = -h(m_{\rho}) - \frac{24\pi}{g_{\rho\pi\pi}^2} - \frac{2q_{\rho}^2}{m_{\rho}}h'(m_{\rho}),& \qquad f_{0} = -\frac{m_{\pi}^2}{\pi} - q_{\rho}^2h(m_{\rho}) - b\frac{m^{2}_{\rho}}{4}, \qquad h(\sqrt{s}) \frac{2}{\pi} 
		\frac{q_{\rm cm}}{\sqrt{s}} \ln\left(\frac{\sqrt{s} + 2q_{\rm cm}}{2m_{\pi}}\right),
	\end{align}
	where $q_{\rho}$ is the center-of-mass momentum at the resonance energy. The 
	curve shown in Fig.~\ref{f:ff} is not a fit but a `prediction' 
	using the values of $m_{\rho}$ and 
	$g_{\rho\pi\pi}$ obtained from the phase shift analysis. We see that this 
	GS model fits our data rather well.

\section{Conclusion}

A first large-volume application of the stochastic LapH method to calculate 
pion-pion scattering on an 
anisotropic lattice has been performed. This results in a good precision for
both the $I=1$ and $I=2$
the phase shift and improved momentum resolution due to the large volume.
While the energies differ significantly from their non-interacting values in 
the $I=1$ $p$-wave irreps, stochastic LapH is sufficiently precise 
to resolve the 
differences for $I=2$ $s$-wave scattering scattering phase shift as well. The 
fewer points below inelastic threshold here are due to the reduced number of 
irreps in which the $s$-wave contributes. A first look at scattering in the 
final isospin 
combination ($I=0$) is underway, but complicated by both `annihilation' 
Wick contractions and the need for a vev subtraction when $\vec{d} = 0$.  

Motivated by this success on the anisotropic lattice, we have started to apply
these techniques to isotropic lattices with lighter pion masses and 
smaller lattice spacings generated by the CLS community effort. This work 
reports on a first preliminary calculation of the $I=1$ $p$-wave scattering 
phase shift on a single $48^2\times 128$ ensemble with 
$m_{\pi} = 280\mathrm{MeV}$ and $a = 0.064\mathrm{fm}$. Using considerably 
fewer Dirac matrix inversions on this ensemble compared with the anisotropic 
one yields results for $m_{\rho}$ and $g_{\rho\pi\pi}$ with comparable 
statistical precision. Apart from the scattering phase shift, we also calculate
the timelike pion form-factor, which is related to the hadronic vacuum 
polarization at low four-momentum transfer. Given the moderate number of 
inversions required, we plan to increase our current level of statistics for 
this form factor by using an additional source time. Furthermore, while we 
average over equivalent orientations of the total momentum $\vec{d}$ in 
calculation of the correlation functions used for $\delta_1$, 
we have not done so for those used for $|F_{\pi}(Q^2)|^2$. Preliminary tests 
indicate that this
averaging has a significant impact on the precision of the form factor and it 
will be performed
on the additional source time, possibly resulting in a significant reduction 
in the statistical errors. Finally, calculations of the scattering phase 
shift and timelike pion form factor on additional CLS ensembles are underway. 

\acknowledgments{BH is supported by Science Foundation Ireland under Grant
	No. 11/RFP/PHY3218. 
CJM acknowledges support from the U.S.~NSF under award PHY-1306805
and through TeraGrid/XSEDE resources provided by
TACC, SDSC, and NICS under grant number TG-MCA07S017. The authors wish to 
acknowledge the DJEI/DES/SFI/HEA Irish Centre for High-End 
Computing (ICHEC) for the provision of computational facilities and support.
The CLS consortium also acknowledges PRACE for awarding access to resource 
FERMI based in Italy at CINECA, Bologna and to resource SuperMUC based in 
Germany at LRZ, Munich.
Furthermore, this work was supported by a grant from the Swiss National
Supercomputing Centre (CSCS) under project ID s384. We are grateful for the
support received by the computer centers.
}

\bibliographystyle{JHEP}   
\bibliography{latticen}

\end{document}